\documentclass[
    ,final            % use final for the camera ready runs
  ]
  {aipproc}

\layoutstyle{8x11single}
\usepackage{amsmath, ulem, natbib, graphicx}

\begin{document}

\title{Angular Momentum Changes Due to Direct Impact Accretion in a 
Simplified Binary System}

\classification{98.10+z, 97.10.Gz, 97.10Kc}
\keywords      {Stars: Binaries: Close, Mass Loss, Direct Impact, 
Angular Momentum Changes}

\author{Jeremy F. Sepinsky}{
  address={Department of Physics, The University of Scranton, Scranton, 
  PA 18510},
}

\author{Bart Willems}{
  address={Department of Physics and Astronomy, Northwestern
  University, 2145 Sheridan Road, Evanston, IL 60208}
}

\author{Vassiliki Kalogera}{
  address={Department of Physics and Astronomy, Northwestern 
  University, 2145 Sheridan Road, Evanston, IL 60208}
}
\author{Frederic A. Rasio}{
  address={Department of Physics and Astronomy, Northwestern
  University, 2145 Sheridan Road, Evanston, IL 60208}  
}

\begin{abstract}

 We model a circular mass-transferring binary system to calculate the 
 exchange of angular momentum between stellar spins and the orbit due to 
 direct impact of the mass transfer stream onto the surface of the 
 accretor.  We simulate mass transfer by calculating the ballistic 
 motion of a point mass ejected from the $L_1$ point of the donor star, 
 conserving the total linear and angular momentum of the system, and 
 treating the stars as uniform density spheres with main sequence radii 
 determined by their masses.  We show that, contrary to previous 
 assumptions in the literature, direct impact does not always act as a 
 sink of orbital angular momentum and may in fact increase it by 
 facilitating the transfer of angular momentum from the spin of the 
 donor to the orbit.  Here, we show an example of the exchange of 
 angular momentum, as well as a measure of the orbital angular momentum 
 changes for a variety of binary star systems with main sequence 
 components.

\end{abstract}

\maketitle

%%%%%%%%%%%%%%%%%%%%%%%%%%%%%%%%%%%%%%%%%%%%
%% MAINMATTER
%%%%%%%%%%%%%%%%%%%%%%%%%%%%%%%%%%%%%%%%%%%%

\section{Introduction}

 In close binary systems, mass transfer is always accompanied by 
 exchange of angular momentum.  In cases where the mass transfer stream 
 directly impacts the accretor, it is commonly assumed that any angular 
 momentum carried by the mass is transferred entirely to the spin of the 
 accretor, thereby removing it from the orbit 
 \citep[e.g.][]{2004MNRAS.350..113M, 2007ApJ...655.1010G, 
 2008A&A...487.1129V}.  Furthermore, many of the same studies assume the 
 angular momentum added to the accretor's spin is identical to the 
 orbital angular momentum the transferred mass had at ejection from the 
 donor, neglecting changes due to gravitational interaction with the 
 binary.

 In order to assess the fate of these systems, accurate calculations of 
 the angular momentum exchange are needed.  For example, calculations of 
 the systemic mass loss in Algol binaries depend strongly upon the 
 rotation rate of the accretor \citep{2008A&A...487.1129V}, while the 
 likelihood that double white dwarfs will be driven to coalesce requires 
 knowledge of the orbital angular moment losses 
 \citep{2007ApJ...655.1010G, 2004MNRAS.350..113M}. In this paper, we 
 briefly present our preliminary results showing a violation of these 
 standard assumptions about orbital angular momentum transport in a 
 simplified binary system.

\section{Calculations}

We consider a circular binary system consisting of two main sequence 
stars with masses $M_D$ and $M_A$, radii ${\cal R}_D$ and ${\cal R}_A$ 
\citep{1996MNRAS.281..219T}, and uniform rotation rates $\Omega_D$ and 
$\Omega_A$, with the subscripts $D$ and $A$ corresponding to the donor 
and accretor, respectively.  We let $\Omega_K$ be the Keplerian orbital 
angular velocity at the periastron of the orbit.  We treat the stars as 
uniform density spheres with inertial constants $k_D=k_A=2/5$, and 
choose the orbital separation $a$ such that the volume equivalent radius 
of the effective Roche lobe \citep{2007ApJ...660.1624S} is equal to 
${\cal R}_D$.

To model the response of the system to mass loss, we eject a single 
particle of mass $M_P \ll M_D,M_A$ from the $L_1$ point of the donor 
star \citep{2007ApJ...660.1624S, 2007ApJ...667.1170S} with a velocity 
equal to the vector sum of the orbital velocity at the $L_1$ point and 
the rotational velocity of the donor star at that point.  The three-body 
system is then evolved via numerical integration of the Newtonian 
equations of motion until $M_P$ impacts the surface of the accretor.  
During both accretion and ejection, the linear and angular momenta of 
the system are conserved.  For details of the calculation, see 
\citet{2010arXiv1005.0625S}.  To model continuous mass transfer in the 
circular orbit, we assume that, for sufficiently small $M_P$, the 
specific angular momentum transferred by a single ejected particle is 
identical to that transferred by a continuous mass transfer stream which 
follows the same trajectory.

\section{Results \& Conclusions}

\begin{figure}
  \label{fig-1}
  \includegraphics[height=.26\textheight]{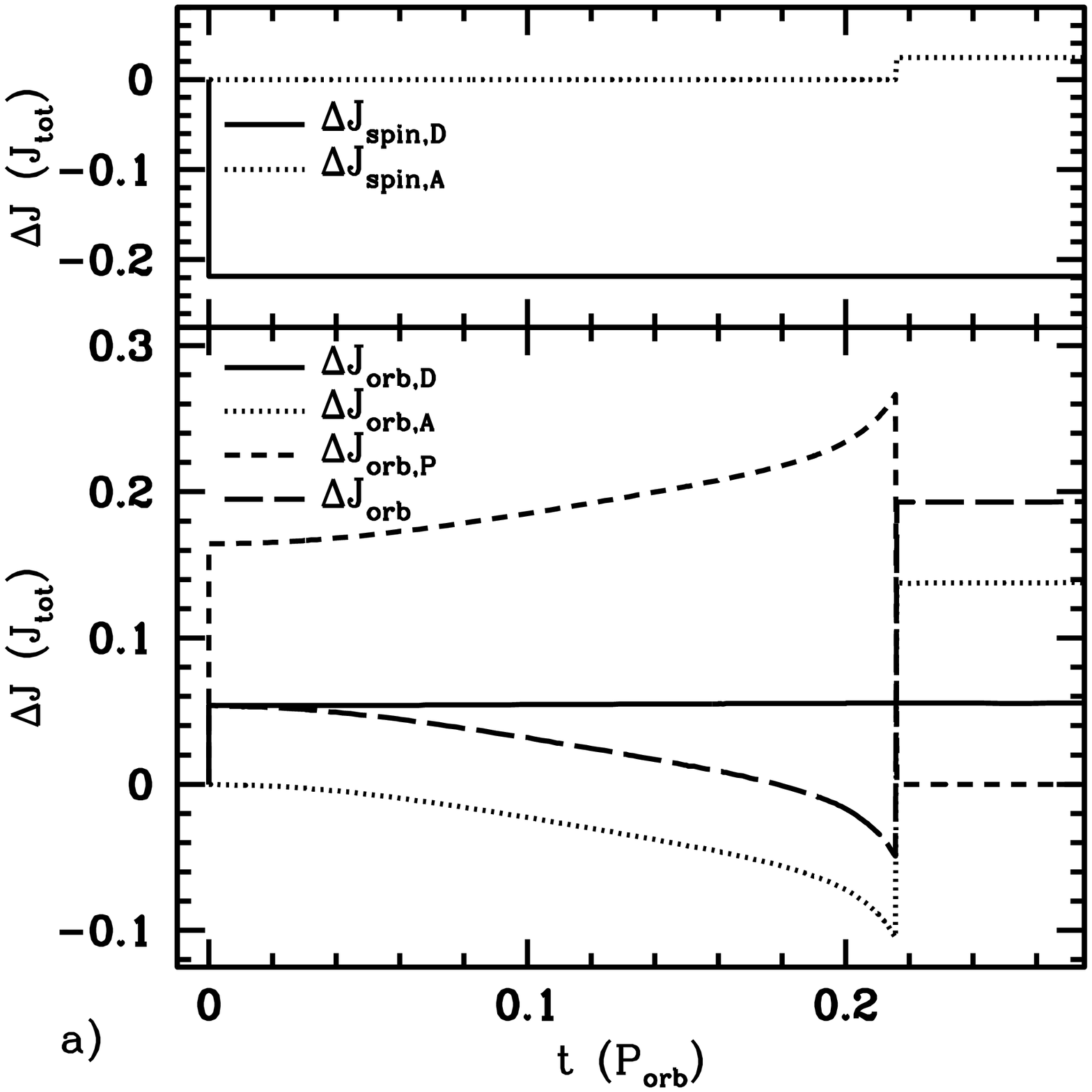}
  \includegraphics[height=.26\textheight]{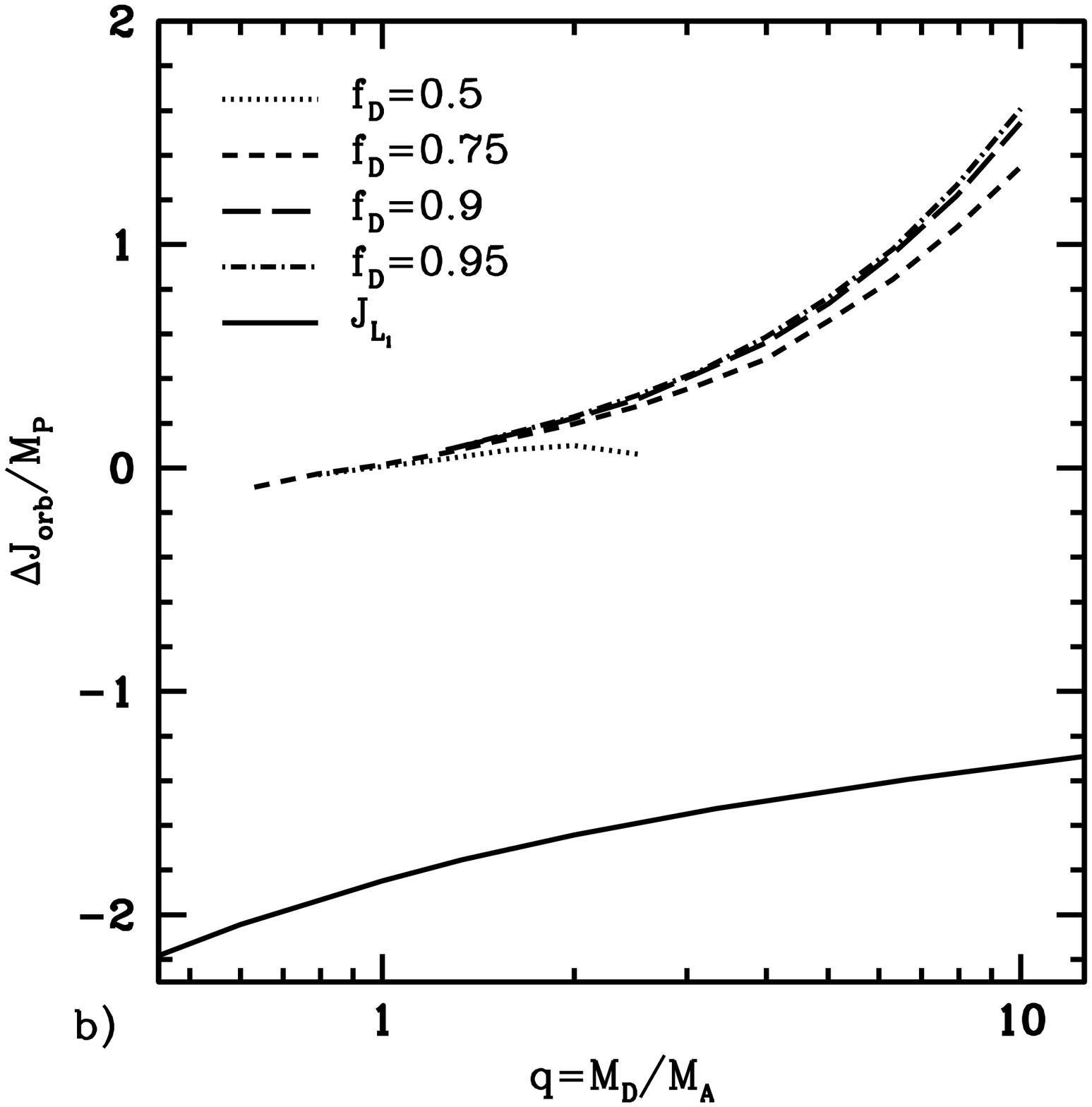}
  \caption{{\bf a)} The change in the spin ({\it top}) and orbital ({\it 
  bottom}) angular momenta between ejection and accretion of a single 
  ballistic particle as a fraction of the total angular momentum for a 
  system with $M_D=10\,M_\odot$, $M_A=1\,M_\odot$, and 
  $\Omega_D=0.9\Omega_K$. {\bf b)} The change in the orbital angular 
  momentum ($\Delta J_{\rm orb}$) of the binary orbit per unit accreted 
  mass ($M_P$) for a system with $\Omega_A=\Omega_K$, 
$M_A=1.0\,_\odot$, 
  and $\Omega_D=f_D\Omega_K$. Note that we can see either an increase 
  ($\Delta J_{\rm orb} > 0$) or a decrease ($\Delta J_{\rm orb} < 0$) 
in 
  the orbital angular momentum of the system depending on the orbital 
  parameters. For comparison, the solid line shows the change in 
orbital 
  angular momentum assuming all of the ejected particle's initial 
  orbital angular momentum is deposited into the spin of the accretor.
  }
\end{figure}

 In Figure~\ref{fig-1}a, we show the change in the spin {\it (top)} and 
 orbital {\it (bottom)} angular momenta for a binary system with the 
 parameters given in the caption.  Mass is ejected at $t=0$, and 
 accretes at $t=0.216\,P_{\rm orb}$ where rapid jumps in the momenta 
 occur due to their conservation at the instantaneous inelastic ejection 
 and accretion.  In this case, nearly all the orbital angular momentum 
 of the particle is transferred to the orbital angular momentum of the 
 accretor, while only a small amount is transferred to its spin.
 
 In Figure \ref{fig-1}b we show the change in the orbital angular 
 momentum per unit $M_P$ for a large number of binary systems. We see 
 that the orbital angular momentum of nearly all the tested systems 
 increases due to direct impact accretion.  During the ejection of 
 $M_P$, conservation of momentum dictates a loss of spin angular 
 momentum of the donor, increasing the orbital angular momentum of the 
 particle.  Upon accretion, a portion of this is added to the accretor's 
 spin while the rest is added to the orbit.  These results are 
 characteristically different from those obtained by the standard 
 assumption of decreasing the orbital angular momentum by the specific 
 angular momentum of the particle at the $L_1$ point ($J_{L_1}$) and 
 adding it entirely to the spin of the accretor.  The solid line in 
 Figure \ref{fig-1}b shows the change in the orbital angular momentum 
 following this perscription. This may have a large effect on the 
 predictions of the survivability of any class of system which undergoes 
 a direct impact mass transfer phase during the course of its evolution, 
 e.g., double white dwarfs, Algol binaries, cataclysmic variables, etc.
 
 In future work, we will explore this intriguing trend in more detail, 
 applying the results to systems such as double white dwarfs where the 
 change in orbital angular momentum is critical to assessing the 
 stability of mass transfer.
  
\bigskip

  The authors would like to thank Christopher Deloye, Paul Groot, and 
  Tom Marsh for useful discussions.

%%%%%%%%%%%%%%%%%%%%%%%%%%%%%%%%%%%%%%%%%%%%%%%%
%% BACKMATTER
%%%%%%%%%%%%%%%%%%%%%%%%%%%%%%%%%%%%%%%%%%%%%%%%

%\begin{theacknowledgments}
%\end{theacknowledgments}

\bibliographystyle{aipproc}   % if natbib is available

%\bibliography{AngMT}
\bibliography{Sepinsky_Mykonos}

\end{document}